\documentclass[draft]{agujournal2019}
\usepackage[english]{babel}
\usepackage{amsmath}
\usepackage{txfonts}
\usepackage{lineno}
\usepackage{setspace}
\usepackage{multirow}
\usepackage{color}
\journalname{JGR: Solid Earth}
\begin{document}
\title{Heat-blanketed convection and its implications for the continental lithosphere}
\authors{K. Vilella\affil{1,2} and F. Deschamps\affil{1}}
\affiliation{1}{Institute of Earth Sciences, Academia Sinica, Taipei, Taiwan}
\affiliation{2}{JSPS International Research Fellow, Hokkaido University, Japan}
\correspondingauthor{Kenny Vilella}{kennyvilella@gmail.com}
\correspondingauthor{Fr\'ed\'eric Deschamps}{frederic@earth.sinica.edu.tw}
\begin{keypoints}
\item We study a mixed heated system where the internal heating is generated only within a horizontal layer close to the surface.
\item  The convective system becomes insensitive to the presence of the heated layer when its thickness is extremely small.
\item When applied to Earth, our results suggest that the presence of continents does not impact significantly Earth's cooling rate.
\end{keypoints}
\begin{abstract}
Earth's continents are characterized by a strong enrichment in long-lived radioactive isotopes.
Recent estimates suggest that they contribute to 33\% of the heat released at the surface of the Earth, while occupying less than 1\% of the mantle.
This distinctive feature has profound implications for the underlying mantle by impacting its thermal structure and heat transfer.
However, the effects of a continental crust enriched in heat-producing elements on the underlying mantle have not yet been systematically investigated.
Here, we conduct a preliminary investigation by considering a simplified convective system consisting in a mixed heated fluid where all the internal heating is concentrated in a top layer of thickness $d_{HL}$ (referred to as ``heat-blanketed convection'').
We perform 24 numerical simulations in 3D Cartesian geometry for four specific set-ups and various values of $d_{HL}$.
Our results suggest that the effects of the heated layer strongly depend on its thickness relative to the thickness of the thermal boundary layer ($\delta_{TBL}$) in the homogeneous heating case ($d_{HL} = 1.0$).
More specifically, for $d_{HL} > \delta_{TBL}$, the effects induced by the heated layer are quite modest, while, for $d_{HL} < \delta_{TBL}$, the properties of the convective system are strongly altered as $d_{HL}$ decreases. 
In particular, the surface heat flux and convective vigour are significantly enhanced for very thin heated layers compared to the case $d_{HL} = 1.0$.
The vertical distribution of heat producing elements may therefore play a key role on mantle dynamics.
For Earth, the presence of continents should however not affect significantly the surface heat flux, and thus the Earth's cooling rate.
\end{abstract}
\section{Introduction}
Thermal evolution of Earth is mainly controlled by the convective transport of heat from its interior to its surface.
This process is traditionally investigated using either numerical simulations of mantle dynamics \cite{Christensen1994, Nakagawa2004, Li2014} or analytical calculations of parametrised convection \cite{Sharpe1978, Honda1995, Grigne2001, Butler2002, Jellinek2015}.
A long-standing challenge for such models is their ability to incorporate self-consistently Earth's continents \cite{Gurnis1988, Zhong1993, Zhong1995, Tackley1998, Tackley2000, Tackley2000b, Lenardic2005, Heron2010, Heron2011, Rolf2011, Rolf2012, Yoshida2013, Jellinek2015}, which are believed to affect the Earth's evolution substantially \cite{Grigne2001, Korenaga2008}.
With the notable exception of \citeA{Cooper2006}, these studies have however neglected the enrichment of continents in long-lived radioactive isotopes.

Recent estimates of the Earth's heat budget suggest that the heating rate in the continental crust is on average about 50 times higher than in the mantle \cite{Jaupart2015}.
Such a large enrichment has important implications for the underlying mantle, as it impacts the heat flow and temperature at the base of the continental lithosphere \cite{Jaupart1998}.
Mantle convection may in turn affect the conditions at the base of the continental lithosphere altering, for instance, their growth due to its thermal sensitivity \cite{Jull2001}.

Here, we perform a preliminary investigation addressing the effects of a heterogeneous source of internal heating on the convective system.
In that aim, we conduct a series of 24 high resolution 3D numerical simulations of ``heat-blanketed convection'', a system where internal heating is generated only within a horizontal layer located close to the top surface.
The thickness of the heated layer is systematically varied in order to quantify its effects on the properties of the system.
In a last section, we discuss the potential implications of our results for the evolution and modeling of continents.

\section{Model of thermal convection for a heterogeneously heated fluid}
\subsection{Physical model}
Thermal convection driven by heterogeneous heating sources is derived from the homogeneous heating case, with which it shares many similarities.
For the sake of clarity, we begin by presenting the more classical homogeneous heating case.
This convective system is composed of a layer of volumetrically heated fluid encased between isothermal top and bottom surfaces.
The top surface is colder than the bottom one such that the base may provide an additional source of heat (bottom heating).
When assuming an isoviscous and incompressible fluid, the system is controlled by two dimensionless numbers, namely the Rayleigh number,
\begin{linenomath*}
\begin{equation}
Ra = \frac{\rho g \alpha \Delta T d^{3}}{ \kappa \eta},
\end{equation}
\end{linenomath*}
and the dimensionless heating rate,
\begin{linenomath*}
\begin{equation}
H= \frac{h d^{2}}{\lambda \Delta T },
\end{equation}
\end{linenomath*}
where $\rho$ is the density, $g$ the gravitational acceleration, $\alpha$ the thermal expansion coefficient, $\Delta T$ the temperature jump across the fluid layer, $d$ the layer thickness, $\eta$ the dynamic viscosity, $h$ the heating rate, $\lambda$ the thermal conductivity and $\kappa = \lambda/\rho C_{p}$ the thermal diffusivity, with $C_{p}$ the specific heat capacity at constant pressure.
The Rayleigh number quantifies the vigour of convection with higher values of $Ra$ implying stronger fluid motions and heat transfer. 
The dimensionless heating rate quantifies the relative importance of internal and bottom heating. 
With increasing heating rate, interior temperatures become hotter, which in turn reduces the heat flux from the core and weakens the power of upwelling plumes.

\begin{figure}
\begin{center}
\includegraphics[width=0.98\textwidth]{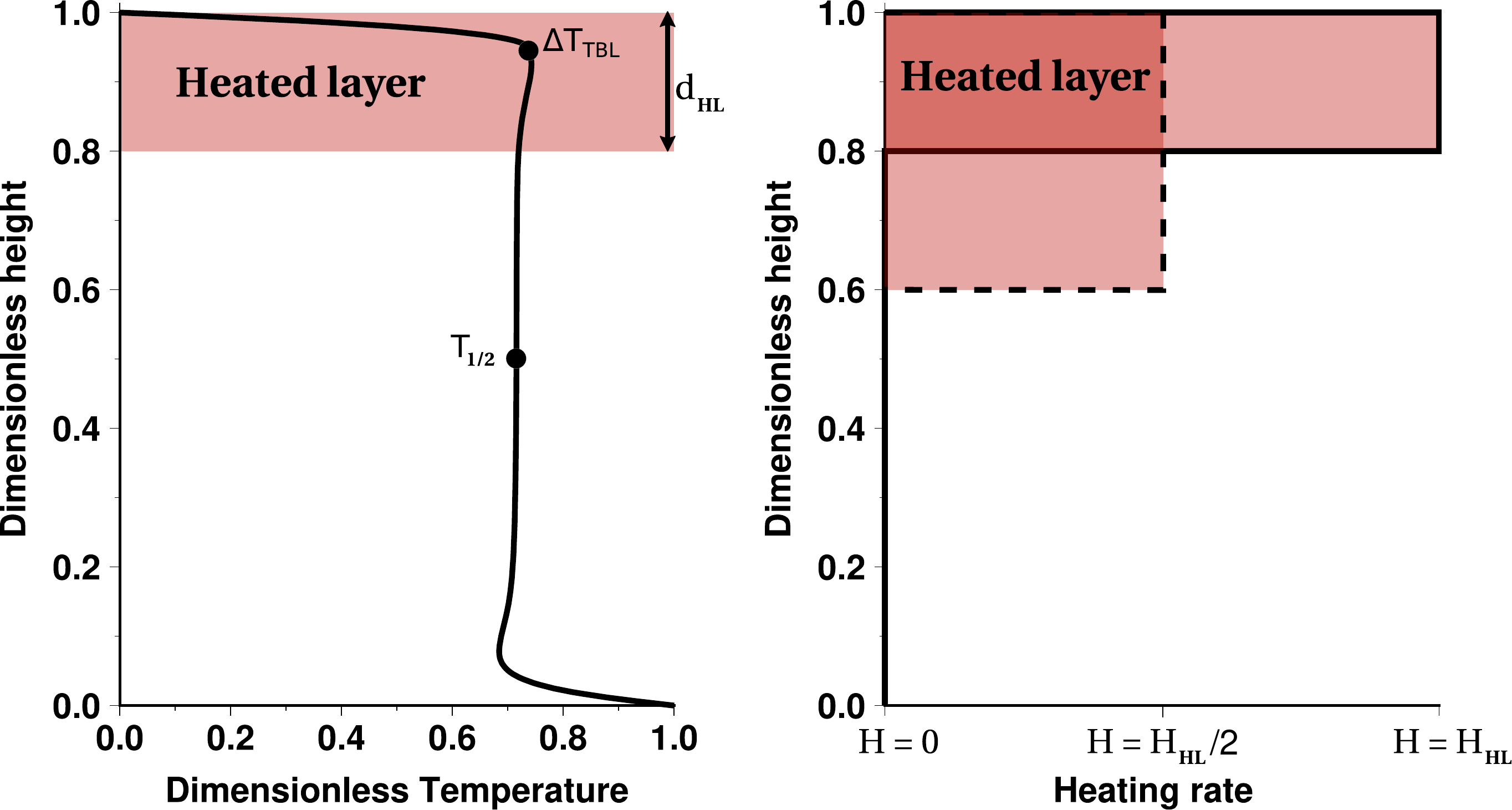}
\end{center}
\caption{\label{Cartoon} Schematic illustration of our physical model. Internal heating (right panel) is only included within a top ``heat blanket'' characterized by a thickness $d_{HL}$ and a heating rate $H_{HL}$ constant with time. The value of the heating rate $H_{HL}$ changes with $d_{HL}$ ($H_{HL} = H / d_{HL}$) in order to keep the same amount of generated heat ($H$). The influence of this layer will be inferred with the temperature profile (left panel) by measuring the temperature jump across the top thermal boundary layer ($\Delta \!T_{TBL}$) and the temperature at mid-depth ($T_{1/2}$).}
\end{figure}

When considering heterogeneous heating, the convective system is identical to the one presented above except that internal heating is allowed to vary in time or/and space.
Here, we seek to understand the fundamental mechanisms induced by heterogeneous heating that are valid independently of the distribution of internal heating. 
As a reference set-up, we therefore concentrate all the internal heating within a horizontal layer that remains stable with time and is located right below the surface, or ``heat blanket'' for short (figure~\ref{Cartoon}).
The horizontal layer has a dimensionless thickness $d_{HL}$, which can be viewed as an additional controlling dimensionless number.
Note that $d_{HL} = 1$ corresponds to the homogeneous case, while $d_{HL} = 0$ corresponds to a Rayleigh-B\'enard system (without internal heating).
We only consider few representative thicknesses, and few representative values of the couple ($Ra$, $H$).
This can be viewed as restrictive, but is necessary, since an exhaustive exploration of the space parameter would require an exaggerated computational time.

When changing the thickness of the heated layer at a given heating rate, one also changes the amount of generated heat.
A different amount of heat is likely to cause important differences in the convective system.
It is therefore more appropriate to compare numerical simulations with the same amount of heat involved.
To do so, we modify the input value for the heating rate ($H_{HL}$) in our numerical simulations, such that the amount of generated heat remains constant when changing the thickness of the heat blanket ($H_{HL} = H / d_{HL}$).
For the description of the model, we mention $H = H_{HL} d_{HL}$ corresponding to the equivalent heating rate in the homogeneous case.
\subsection{Numerical model \label{NumModel}}
The numerical simulations of heat-blanketed convection are performed in three dimensional Cartesian geometry using the code StagYY \cite{Tackley2008}, which solves the dimensionless conservation equations of mass, momentum and energy.
Cartesian geometry may not be relevant for planetary mantles, but it allows to capture the effects of heterogeneous heating in a straightforward way.
The top and bottom surfaces are isothermal and the mechanical boundary conditions are free slip, whereas the lateral boundary conditions are reflecting.
All fluid properties are constant, except for density that depends on temperature in the buoyancy force \cite<usually referred to as the>[approximation]{Boussinesq1903}.
The initial temperature condition within the whole box is constant with a dimensionless value between 0.5 and 1.0 depending on the conditions considered.
Note that random, small amplitude (0.01 or 0.001), perturbations are superimposed to this constant temperature in order to trigger convection.
We stop the numerical simulations when a statistical steady state is reached.
In practice, we determine the steady state as the stage for which both the volumetric average temperature and the surface heat flux are constant (their fluctuations are zero) when averaged over several overturn times.

For each couple ($Ra$, $H$) considered, we run one numerical simulation with a homogeneous heating ($d_{HL} = 1$) along with 4 to 8 additional simulations conducted for different values of $d_{HL}$ (listed in table~\ref{run}).
The selected aspect ratio and grid resolution (table~\ref{run}) guarantee both the development of a large number of convective currents (with a size that does not depend on the aspect ratio), and a good resolution in the thermal boundary layers (TBL) and convective currents (see Supplementary Material for more details).
Moreover, this set of numerical simulations provides a good coverage for the possible values of the dimensionless numbers relevant to planetary bodies.
\section{Heat-blanketed convection}
The effects of the heat blanket on the convective system can be expressed by different processes, including changes in the temperature distribution and in the shape of convective currents.
We therefore separate this exploration in three parts:
(i) influence of the heat blanket thickness $d_{HL}$ on the global characteristics of the convective system;
(ii) variations in temperature profiles;
and (iii) impacts on the flow pattern.
\subsection{Global characteristics}
Characterizing thermal convection is challenging because the main properties of the system strongly vary both in space and time.
A classical approach to overcome this issue is to focus on parameters that are spatially and/or temporally averaged.
In particular, from a theoretical point of view, the convective system is often characterized using the horizontally and temporally averaged values of the surface heat flux ($\phi$), temperature jump across the top thermal boundary layer ($\Delta \!T_{TBL}$), and temperature at mid-depth ($T_{1/2}$), these properties providing a good description of the thermal state.
In addition to the thermal state, it is useful to quantify the vigour of convection.
For instance, the convective vigour can be quantified using the temporally and volume averaged root mean square velocity ($V_{rms}$).
We also report the temporally and horizontally averaged surface velocity ($V_{h}$), which has the advantage to be easily measurable on Earth \cite<e.g.,>[]{Sella2002}.
These properties are reported in table~\ref{run} for all the numerical simulations.
Note that $\Delta \!T_{TBL}$ is here measured from the ``hot'' temperature profile composed of the hottest temperature at every depth (a discussion on the methods to define the thermal boundary layer is provided in Supplementary Material).
For the sake of simplicity, we will focus the description of the results only on the case conducted at $Ra = 10^{6}$ with $H = 20$, for which we have investigated a larger set of $d_{HL}$ values.
We have selected this specific case because the system is vigourous enough to represent Earth-like convection, while the required computational time remains reasonable.
Conclusions for this specific case are however fully consistent with results obtained for the three other cases (see Supplementary Figure S3).

The results for the five properties investigated are plotted in figure~\ref{HL_effects} as a function of $d_{HL}$.
For $d_{HL} = 0$, we report the results obtained by \citeA{Vilella2018a} for pure bottom heating ($H = 0$).
Figure~\ref{HL_effects} shows that the properties are changing continuously from values obtained without internal heating ($d_{HL} = 0$) to values obtained for homogeneous heating ($d_{HL} = 1$).
This change is however not linear and is different for each property, so that a separate description is required.
From $d_{HL} = 1$ to $0.1$, the surface heat flux is slightly decreasing ($\sim 3\%$) and then increases very sharply ($\sim 40\%$) until $d_{HL} = 0$.
Interestingly, we note that the transition between these two different behaviours occurs when the thickness of the heat blanket becomes smaller than the thickness of the top thermal boundary layer ($\delta_{TBL}$) measured in the homogeneous case (blue shaded area).
This observation is not surprising given that in internally heated systems the top thermal boundary layer controls the dynamics of the system.
More specifically, when $d_{HL} > \delta_{TBL}$ the top thermal boundary layer is fully internally heated, which, following the theoretical framework of \citeA{Vilella2017}, implies a direct relationship between the thermal structure of the top thermal boundary layer, the surface heat flux and the Rayleigh number.
Because the Rayleigh number is here constant, the surface heat flux and $\Delta \!T_{TBL}$ remain constant (figure~\ref{HL_effects}a,b), even if the interior temperature $T_{1/2}$ sharply decreases.
The interior temperature $T_{1/2}$ is decreasing with $d_{HL}$ simply because a decreasing portion of the convective interior is being heated.
For $d_{HL} < \delta_{TBL}$, the top thermal boundary layer is no longer entirely heated so that it does not fully control the dynamics of the system.
In particular, the thermal structure and heat transfer of the thermal boundary layer are now affected by hot plumes originating from the opposite thermal boundary layer.
As for the surface heat flux, the temperature of the system is varying sharply until reaching the values obtained without internal heating.
Surprisingly, $\Delta \!T_{TBL}$ and $T_{1/2}$ reach a peak when $d_{HL} \approx \delta_{TBL}$, a behaviour that we discuss in section~\ref{secTProfil}.
Finally, the velocities $V_{h}$ and $V_{RMS}$, characterizing the vigour of convection, are highly correlated with one another, and are further correlated to the surface heat flux.
This is not surprising since higher heat flux requires a more vigourous convection to transport the corresponding amount of heat.
Interestingly, velocities are multiplied by a factor $\sim$5 when $d_{HL}$ decreases from 0.15 to 0, indicating that the enrichment of the internal heating in a thin layer strongly increases the convective vigour.
\subsection{Temperature profiles \label{secTProfil}}
Horizontally averaged temperature profiles for representative cases are plotted in figure~\ref{TProfil} (for other cases see Supplementary Figure S4).
As suggested in figure~\ref{HL_effects}b, the system is becoming colder as $d_{HL}$ decreases (figure~\ref{TProfil}a,b).
We also observe that the thickness of the top thermal boundary layer is decreasing with decreasing $d_{HL}$.
This can be easily explained by noting that the surface heat flux does not vary significantly, while a simple dimensional analysis indicates 
\begin{linenomath*}
\begin{equation}
\phi \sim k \frac{T_{1/2}}{\delta_{TBL}},
\label{HeatFlux}
\end{equation}
\end{linenomath*}
implying that a lower $T_{1/2}$ should be balanced by a lower $\delta_{TBL}$.
For $d_{HL} < \delta_{TBL}$, the top thermal boundary layer is characterized by two trends with different temperature variations with depth (figure~\ref{TProfil}a for $d_{HL} = 0.1$ and $d_{HL} = 0.05$).
Within the heated layer, the thermal structure is equivalent to the one obtained for larger $d_{HL}$, while below the heat blanket the thermal boundary layer tends to get closer to the thermal structure obtained in the case without internal heating.
This behaviour indicates that the thickness of the heated layer has to be extremely thin in order to have no effects on the thermal structure of the system.
For $d_{HL} > \delta_{TBL}$, the thermal structure of the top thermal boundary layer only slightly changes with variations of $d_{HL}$.
As a consequence, the temperature is clearly hotter within the heat blanket than below (figure~\ref{TProfil}b for $d_{HL} = 0.3$).
For $d_{HL} \approx \delta_{TBL}$, we can see the formation of a peaked temperature at a depth corresponding to the base of the heated layer.
As previously shown in figure~\ref{HL_effects}b, the interior temperature is slightly hotter in that case than for larger or lower values of $d_{HL}$ (figure~\ref{TProfil}b for $d_{HL} = 0.05$ compared to $d_{HL} = 0.02$ and $0.1$).

In order to gain insight on the generation of instabilities, we have also plotted in figure~\ref{TProfil} the ``hot'' temperature profiles composed of the hottest temperature at every depth.
This profile is particularly appropriate to assess the stability of the top thermal boundary layer \cite{Vilella2017}.
Figure~\ref{TProfil}d shows that the ``hot'' temperature profile is also characterized by a peaked temperature, inducing a thicker, thus more stable, thermal boundary layer.
As a consequence, the generation of cold instabilities should be more difficult, which in turn should reduce their ability to cool the convective interior.
This may explain the increased interior temperature compared to cases with slightly larger or lower $d_{HL}$ values.
Another interesting result is that, for $d_{HL}$ significantly lower than $\delta_{TBL}$ (figure~\ref{TProfil}c for $d_{HL} = 0.1$ and $d_{HL} = 0.05$), the temperature profile does not change when $d_{HL}$ is further reduced.
We therefore expect that for cases with a very thin heated layer, further reducing $d_{HL}$ does not alter the generation of cold downwellings (figure~\ref{TProfil}c), but only modifies the thermal structure of the system (figure~\ref{TProfil}a).
\subsection{Planform of convection\label{PlanformSec}}
\begin{figure}
\begin{center}
\includegraphics[width=0.95\textwidth]{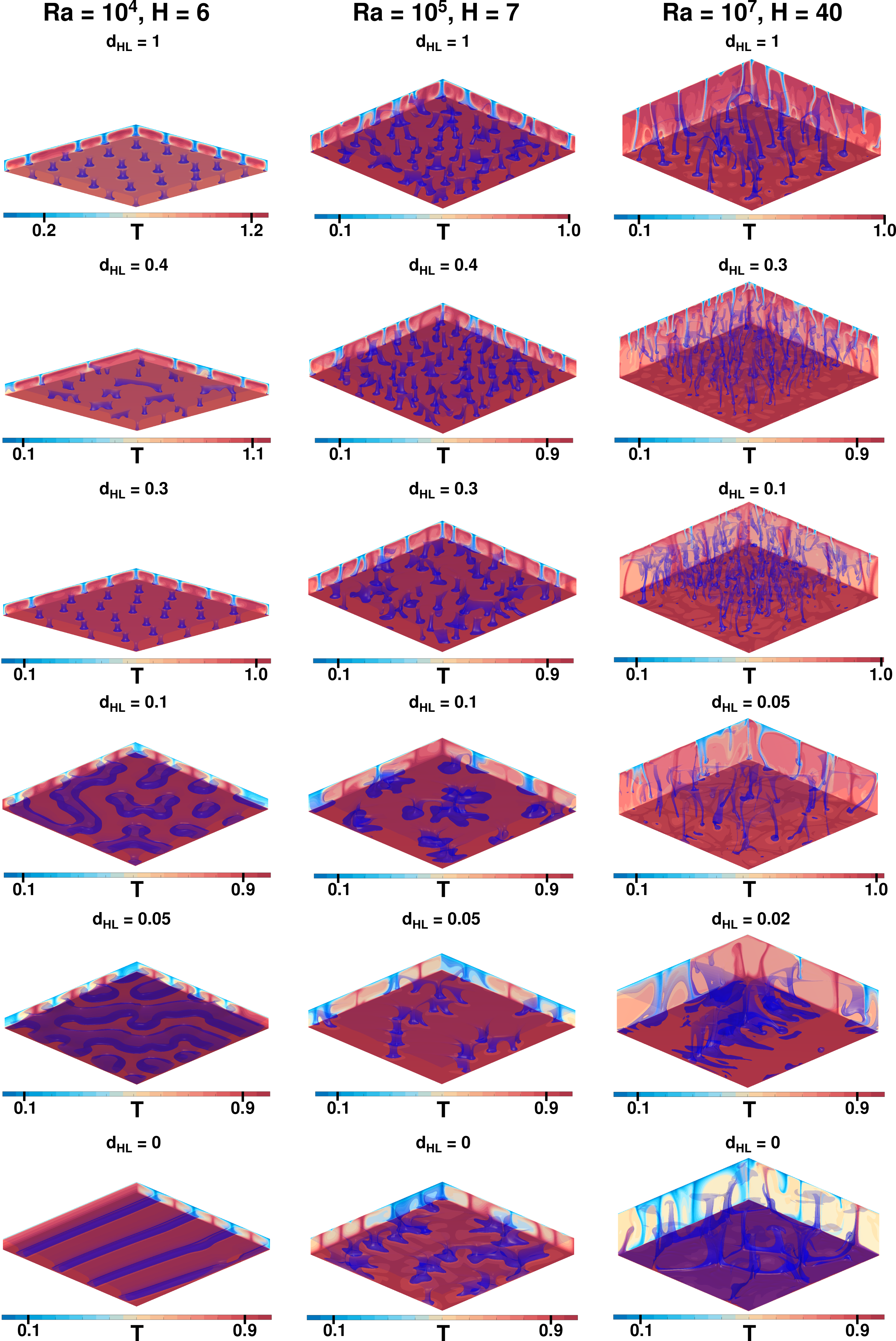}
\end{center}
\caption{\label{Isosurface} Isosurfaces of the cold downwellings for some cases described in table~\ref{run}. The colour scale changes for each simulation in order to enhance the visibility of the convective structure. The domain aspect ratio of the temperature fields represented is kept constant at a given Rayleigh number corresponding to the lower aspect ratio available (table~\ref{run}).
}
\end{figure}

\begin{figure}
\begin{center}
\includegraphics[width=0.95\textwidth]{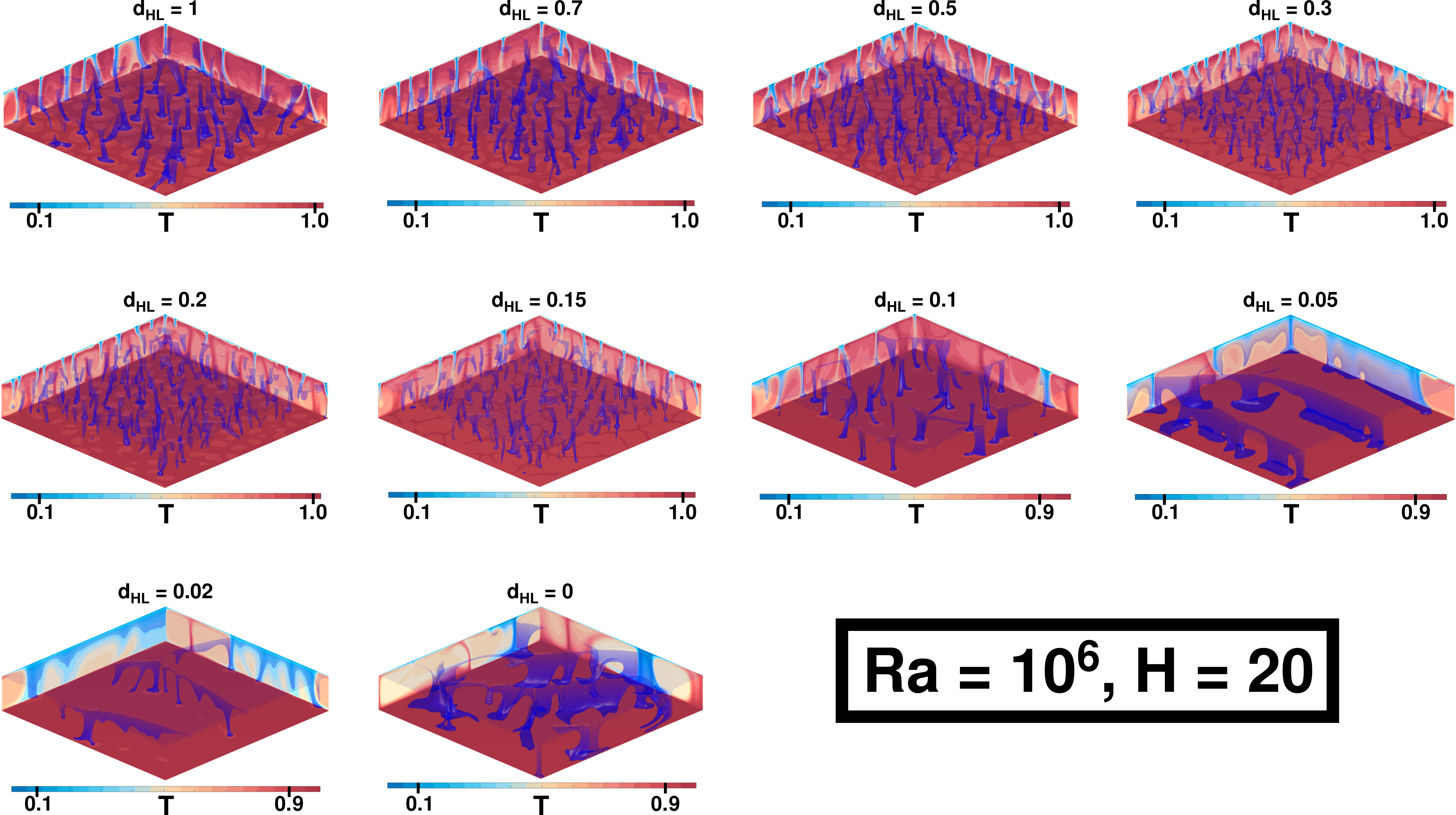}
\end{center}
\caption{\label{Isosurface_2} Isosurfaces of the cold downwellings for cases with $Ra = 10^{6}$ and $H = 20$ described in table~\ref{run}. The colour scale changes for each simulation in order to enhance the visibility of the convective structure. The domain aspect ratio of the temperature fields represented is kept constant to the lower aspect ratio available (table~\ref{run}).
}
\end{figure}

Previous observations have provided critical information on convective motions.
For instance, the change of convective vigour may be deduced from the thickness of the heated layer.
These inferences are however based on implicit arguments.
In order to reach more robust conclusions, one may combine previous observations with the description of convection planforms reported in figures~\ref{Planform} to~\ref{Isosurface_2}.
The systematic description of convection planforms has been conducted for Rayleigh-B\'enard systems considering an isoviscous \cite{Busse1971, Whitehead1978, Houseman1988, Travis1990} or strongly temperature dependent viscosity fluid \cite{White1988, Christensen1991}, and for purely internally heated systems \cite{Houseman1988, Travis1990, Limare2015, Vilella2018b}.
Such a study is however missing for mixed heating convection.
By analogy with these extensive data sets, one can however draw general statements on the pattern obtained in this peculiar system.

In the homogeneous heating case ($d_{HL}=1.0$), the planforms are composed of focused cold downwellings with cylindrical shape at low $Ra$ and more irregular elongated shape at larger $Ra$ encased in an almost isothermal background.
Hot upwellings are diffused and can be considered as a diffuse non-buoyant return flow.
These characteristics indicate the prevalence of internal heating over bottom heating.
Furthermore, convection planforms reported for purely internally heated convection \cite{Vilella2018b} are qualitatively similar to the ones observed here.
Interestingly, buoyant hot upwellings appear when $d_{HL}$ is decreased suggesting that the importance of internal heating relative to bottom heating decreases, a trend supported by the changes in Urey ratio as a function of $d_{HL}$ (table~\ref{run}).
Moreover, the strength and number of hot upwellings increase with decreasing $d_{HL}$.
For $d_{HL} << \delta_{TBL}$, the convection planform even becomes similar to the case without internal heating ($d_{HL}=0$) with cold downwellings and hot upwellings of equal strength and number.
For such cases, $\Delta \!T_{TBL}$ does not change with varying $d_{HL}$ (figures~\ref{HL_effects} and S3) supporting our choice of using the ``hot'' temperature profile to assess the generation of cold downwellings by the top TBL.
Furthermore, it is worth noting that, for $Ra = 10^7$, the case with the lowest $d_{HL}$ is only slightly lower than $\delta_{TBL}$.
As a result, the planform of convection still exhibits some variations between the cases $d_{HL} = 0.02$ and $d_{HL} = 0$.
This is unlikely caused by the higher Rayleigh number, but simply by the value of $d_{HL}$.
For $d_{HL}$ slightly lower than $\delta_{TBL}$ (for instance $Ra = 10^7$ and $d_{HL} = 0.02$), the hot upwellings and cold downwellings are forming sheet-like structures extending over a large spatial scale.
The width of these convective structures are similar to the cases without internal heating, while the typical distance between two structures is increased.
It may be possible that the thickness of the heat blanket introduces a new, thinner, spatial scale in the convective motions that strongly influences the pattern of convection.
\section{Application to Earth's continents}
Our numerical simulations have shown the potentially important effects of the vertical distribution of internal heating on the convective system.
However, these effects highly depend on the relative thickness of the heated layer ($d_{HL}$) compared to the thickness of the thermal boundary layer ($\delta_{TBL}$) in the homogeneous heating case ($d_{HL} = 1.0$).
More precisely, for $d_{HL} > \delta_{TBL}$, only the interior temperature is affected by variations of $d_{HL}$ (figure~\ref{TProfil}).
By contrast, for $d_{HL} < \delta_{TBL}$, all the properties of the convective system are changing importantly in response to potentially small changes of $d_{HL}$.

Interestingly, our model may be particularly appropriate to model Earth's continents, since they are characterized by a strong enrichment in heat producing elements.
Indeed, continents occupy less than 1\% of the Earth's mantle while contributing to 33\% of the heat produced by the decay of long-lived radioactive elements \cite{Jaupart2007, Jaupart2015}.
The application of our numerical results to Earth's continents requires first to identify the appropriate values for $d_{HL}$ and $\delta_{TBL}$.
The estimate of $d_{HL}$ is straightforward because it simply corresponds to the thickness of the continental crust.
This thickness slightly varies across the globe with an average value of $\sim$40~km and a maximum thickness of $\sim$70~km \cite{Mooney1998, Pasyanos2014}.
Taking a mantle thickness of 2890~km, we obtain dimensionless thicknesses of $d_{HL} \approx 0.013$ on average and $d_{HL} \approx 0.024$ at a maximum so that the cases with $d_{HL} = 0.02$ may be appropriate to represent today's Earth continents.
Determining the thickness of the thermal boundary layer, or equivalently the Rayleigh number, of the mantle is more challenging.
Different attempts to measure $\delta_{TBL}$ on Earth suggest a thickness of about 200~km \cite<e.g.,>[]{Lee2005}.
This corresponds to $\delta_{TBL} \approx 0.07$ and is compatible with values obtained for $Ra = 10^7$.
In the past, it is likely that $\delta_{TBL}$ and $d_{HL}$ were thinner, due to more vigourous convection, so that a larger $Ra$ should be more appropriate for early Earth.

The application of our results to Earth's continents is however not straightforward.
In particular, the models of convection presented here are clearly simplified compared to the Earth's mantle dynamics.
It is therefore crucial to discuss the applicability of our models before applying our results to Earth's continents.
\subsection{Potential applicability to Earth's continents \label{secApplicability}}
The purpose of this work is to identify the effects of a top layer enriched in heat-producing elements on the dynamics of the system.
As such, many complexities of planetary mantles have been neglected or simplified in order to avoid any competitive effects.
The advantage of this approach is to better understand the effects of different ingredients taken separately.
The shortcoming, however, is the difficulty of predicting the effects of a single ingredient on the actual system, as the relative importance between different ingredients is difficult to assess.
In other words, features not included in our model may potentially erase or even reverse some of our conclusions.
It is therefore important to evaluate the potential impact of the main ingredients that are not included in our model.

A simplification of our model is to assume an isoviscous fluid rather than a complex rheology typical of mantle rocks.
Rheology is a long-standing issue in models of planetary mantles.
Traditionally, Newtonian rheologies with temperature and pressure dependent viscosity is considered for the bulk mantle, while rheology with plastic yielding have been used at the surface to mimic plate-like behaviour \cite{Rolf2011, Rolf2012, Yoshida2013}.
The introduction of strongly temperature-dependent viscosity induces the formation of a stagnant lid at the top surface, where fluid motions are inhibited and conductive heat transport dominates.
In addition, because heat is less easily extracted, the interior temperature is larger than in the isoviscous case.
The flow pattern is dominated by plumes that are stopped beneath the stagnant lid.
These properties are observed for both bottom heated and mixed-heated systems \cite{Stein2013b}.
Note that the mobility of the lid decreases with increasing heating rate, that is the lid gets stiffer.
The stagnant lid is often considered as an upward extension of the top TBL.
In a planet whose mantle is animated by stagnant lid convection, we therefore expect that concentrating heat in a thin top layer would have effects similar to cases with $d_{HL} < \delta_{TBL}$.
A strong and thick stagnant lid, however, is not observed on Earth.
Plastic yielding is therefore introduced as a weakening mechanism to allow for the break-up of this lid, leading to a so-called mobile-lid regime.
While a stagnant lid is not present, the interior temperature and heat flux remain larger and lower, respectively, than in the isoviscous case, leading to thicker TBLs.
Overall, the introduction of a more realistic rheology should impact importantly the convection planform \cite{White1988}, while increasing the interior temperature and thickness of the top TBL, such that the case $d_{HL} < \delta_{TBL}$ would be reached more easily.

A second important difference between our model and Earth's continents is that our heat-blanket covers the whole system, while Earth's continents cover only 30\%-40\% of Earth's surface \cite{Taylor1995}.
The effects of continent coverage on the convective system have been investigated with both numerical simulations \cite{Lenardic2005, Cooper2006, Jellinek2009, Coltice2014, Whitehead2015} and laboratory experiments \cite{Guillou1995, Lenardic2005, Jellinek2009}.
A major finding is the existence of two different regimes depending on the continent coverage \cite{Lenardic2005}.
For a surface area covered by continents lower than about 40\%, the heat flux remains high with plate-like motions at the top surface, whereas, for a larger continent coverage, a stagnant lid appears at the top, reducing significantly the surface heat flux.
Note, however, that this transition occurs at greater continent coverage with decreasing Rayleigh number and even disappears for low $Ra$. 
When applied to Earth, one may expect a very moderate effect of continent coverage on the surface heat flux, as Earth should remain throughout its evolution in the high heat flux regime.
Nevertheless, partial coverage of continents should still impact mantle dynamics by inducing a large scale motion in the system \cite{Guillou1995} and by increasing the production rate of sea floor \cite{Coltice2014}.
Moreover, continental drift may affect the convective motions in the mantle by focusing cold downwellings on the edge of continents and disrupting the convective cells \cite{Whitehead2015}.

It is also important to note that the heat blanket in our numerical model is a part of the convective system, while a long-standing hypothesis is that continents do not participate in convection.
An argument supporting this hypothesis is the low amount of material exchange between continents and the mantle.
This formulation may however be misleading because, although continents remain mostly separated from the mantle, several evidences, such as the small heat flux inferred below continents compared to the estimated heat flux below the oceanic lithosphere \cite{McKenzie1988,Putirka2007, Jaupart2011, Jaupart2016}, show that continents have an impact on the thermal structure of the mantle and thus its dynamics.
As such, the incorporation of the heat blanket within the convective system may be a reasonable assumption.

A last difference between our numerical models and Earth is the presence of internal heating not only in the enriched layer but also in the bulk system.
To overcome this issue, one may consider an idealized version of Earth's evolution.
At first, the mantle should be characterized with a homogeneous distribution of internal heating, while the subsequent evolution and chemical differentiation should induce the creation and thickening of a top layer enriched in heat producing elements.
At one point, the enriched layer reaches a critical thickness $d_{HL,cr}$ for which all the long-lived radioactive elements have been extracted from the mantle.
For Earth, $d_{HL,cr}$ should be around $0.06$, at first order.
One may further note that the initial and final stages of this evolution can be represented by our numerical models.
We may therefore use these two end-member cases to constrain the overall effect of continents on the convective system.
More specifically, considering the representative case $Ra = 10^7$, we can compare the results from the initial stage ($d_{HL} = 1.0$) and final stage ($d_{HL} = 0.06$, which can be approximated by our case at $d_{HL} = 0.05$) using results reported in table~\ref{run} and Supplementary Figure S3.
These results indicate that, except for the interior temperature, properties are only slightly impacted by the presence of the heated layer.
In particular, the surface heat flux should not be impacted by the presence of the heated layer, which is likely to be valid for more complex systems, as shown by results of \citeA{Cooper2006}.

\subsection{Are continents preventing Earth cooling?}
It has been suggested that the heat produced by the radioactive isotopes in the continental crust does not participate in convection, and therefore should not be included in the long-term evolution of the Earth's mantle.
An interesting by-product of this hypothesis is that enriching the continental crust in long-lived radioactive isotopes allows removing heat from the Earth's mantle, which in turn induces a lower cooling rate \cite{Grigne2001}.
This has been used, for instance, to explain the thermal evolution of Earth \cite<e.g.,>[]{Jellinek2015}.

An evidence supporting this scenario is the low heat flux beneath the continental lithosphere compared to that beneath the oceanic lithosphere \cite<$\approx65$~mW~m$^{-2}$ and $\approx100$~mW~m$^{-2}$, respectively, following>[and reference therein]{Jaupart2011, Jaupart2016}.
However, the differences between the oceanic and continental lithospheres are such that a direct comparison between these two regions may not be relevant.
For instance, a specificity of the oceanic lithosphere is that heat is dominantly transported by volcanism \cite{Jaupart2015}, which stands as a much more efficient way to transport heat.
Overall, our results suggest that the continental crust is thick enough to affect significantly mantle dynamics.
It is therefore inappropriate to consider the heat produced in the continental crust as separated from the mantle.
As discussed in section~\ref{secApplicability}, and in agreement with \citeA{Cooper2006}, the surface heat flux, and thus the cooling rate, should not be significantly affected by the partitioning of heat in the heated layer.

It is important to note that our results are valid as long as solid-state thermal convection and thermal conduction are the dominant heat transfer processes.
In particular, volcanism is another mode of heat transfer that is not considered in our models and that should change importantly the conclusions we draw.
This, when applying to the Earth's case, explains why our results should not be applied to the oceanic lithosphere, while they may be relevant for describing continents since they are not dominated by volcanism.
\subsection{Control on the thickness of the continental crust}
Observations have shown that stable continents are characterized by a very homogeneous thickness of 35--45~km \cite{Mooney1998, Pasyanos2014}.
At first glance, this homogeneity seems inconsistent with the complexity of Earth's dynamics and the large compositional heterogeneities of the Earth's mantle.
To overcome this issue, the presence of some mechanisms controlling the thickness of the continental crust is traditionally invoked.
In particular, it has been proposed that the pressure and temperature conditions play a key role in the production of the continental crust \cite{Jull2001}.
Below a certain depth, the conditions are no longer suitable for the production of continental crust, which sets an upper limit for the thickness of stable continents.
Considering the simplifications of our model, it is not possible to draw clear and safe conclusions on the potential effects of the heated layer on the thickness of Earth's continents.

Alternatively, an interesting exercise is to study a different rocky planet, for instance a Mars-like planet.
In that case, due to the lower gravitational acceleration, the limiting depth for the production of crust may be significantly deeper.
Interestingly, following the exact crustal thickness, it is possible to reach the regime where $\delta_{TBL} \approx d_{HL}$.
Assuming that the increase of temperature reported in that regime (figure~\ref{HL_effects}) is still present in a more realistic system, one may expect an increase in the crustal production rate.
After this episode of enhanced crustal production, the temperature should decrease while the thickness of the thermal boundary layer should increase (figure~\ref{TProfil}c,d), both effects making the generation of volcanism more difficult.
It is therefore possible that on Mars-like planet the crustal thickness is bound to be slightly larger than $\delta_{TBL}$.
This specific condition may be characterized by a slightly lower surface heat flux (figure~\ref{HL_effects}a) as well as a more sluggish convection (figure~\ref{HL_effects}c), which is compatible with our understanding of Mars dynamics.

\subsection{Are continents acting as insulators?}
It is generally thought that continents act as thermal insulators, reducing the surface heat flux.
This idea originates from the small heat flux inferred below continents compared to below the oceanic lithosphere \cite{McKenzie1988,Putirka2007, Jaupart2011, Jaupart2016}. 
This observation has led some authors to model continents by prescribing an insulator layer at the top surface \cite{Heron2010, Heron2011}.
Alternatively, other authors prescribed a large viscosity jump within continents \cite{Gurnis1988, Zhong1993, Lenardic2005, Lenardic2011, Rolf2011, Rolf2012, Cooper2013} inducing a thicker thermal boundary layer, in agreement with the higher thickness of the continental lithosphere compared to the oceanic lithosphere, which in turn causes an insulation effect.
These models, however, do not include the enrichment in heat producing elements within continents.

Our results suggest that enriching a superficial layer in radioactive isotopes should induce a modest variation of the surface heat flux, which may be viewed as inconsistent with observations.
In that case, however, the surface heat flux does not necessarily reflect the heat flux below the heated layer, since a large amount of heat is produced within the heated layer.
Actually, the heat flux below the heat blanket seems to be minimum when $d_{HL} \approx \delta_{TBL}$, while remaining low for cases appropriate for the Earth's mantle.
Therefore, it appears that the enrichment of the continental crust in radioactive isotopes may be partly responsible for the small heat flux inferred below continents.

\section{Conclusion}
\begin{figure}
\begin{center}
\includegraphics[width=0.9\textwidth]{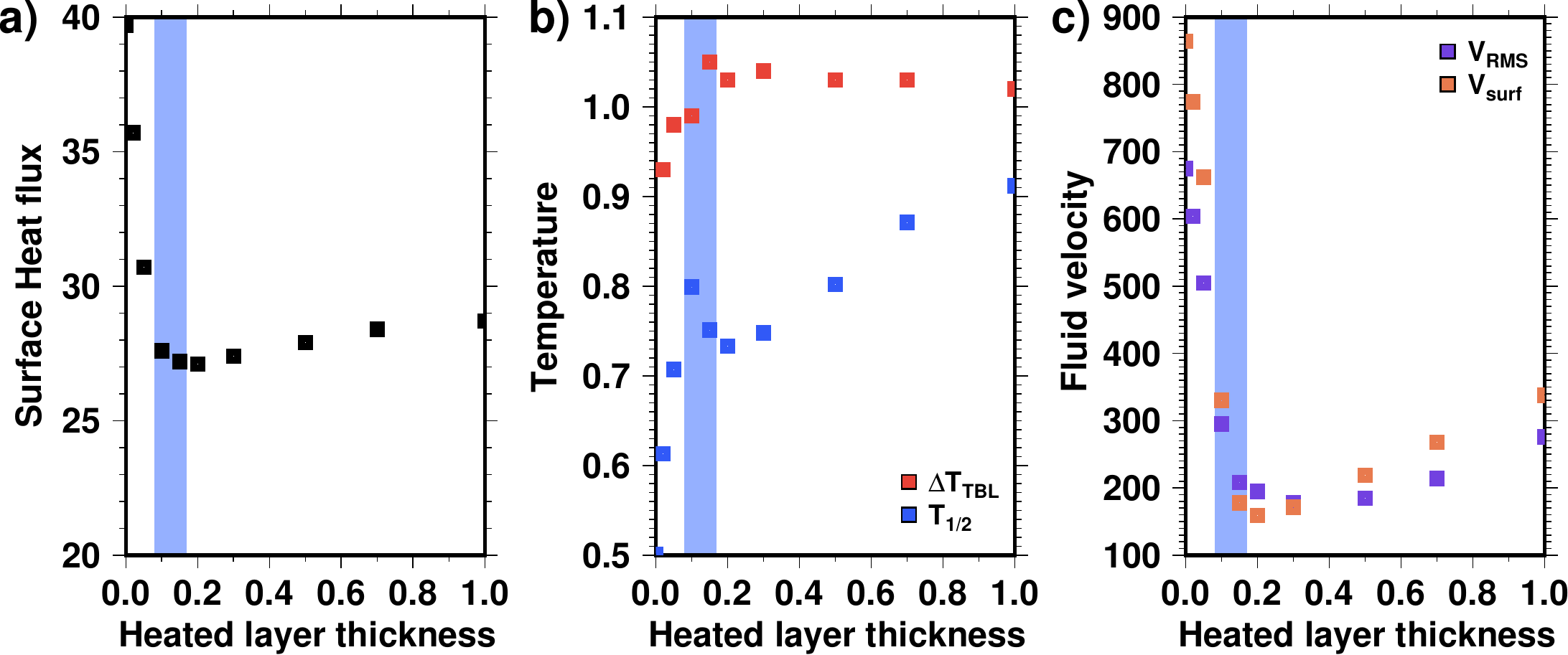}
\end{center}
\caption{\label{HL_effects} Variations of the dimensionless (a) surface heat flux, (b) temperature jump across the top thermal boundary layer ($\Delta \!T_{TBL}$) and average temperature at mid-depth ($T_{1/2}$), (c) average surface velocity ($V_{h}$) and volume average root mean square velocity ($V_{rms}$) as a function of the dimensionless thickness of the heat blanket ($d_{HL}$). The numerical simulations are conducted for $Ra = 10^{6}$ and $H = 20$. The blue shaded area corresponds to the typical values for the thickness of the top thermal boundary layer (see Supplementary Material for more details) in the case where internal heating is homogeneous ($d_{HL} = 1$). The case without internal heating ($d_{HL} = 0$) is from \citeA{Vilella2018a}. Note that for this case the dimensionless surface heat flux is $\sim$19.7. However, in order to achieve a meaningful comparison with cases including internal heating, we add to this number the amount of internal heating included in the other cases ($H = 20$).}
\end{figure}

\begin{figure}
\begin{center}
\includegraphics[width=0.85\textwidth]{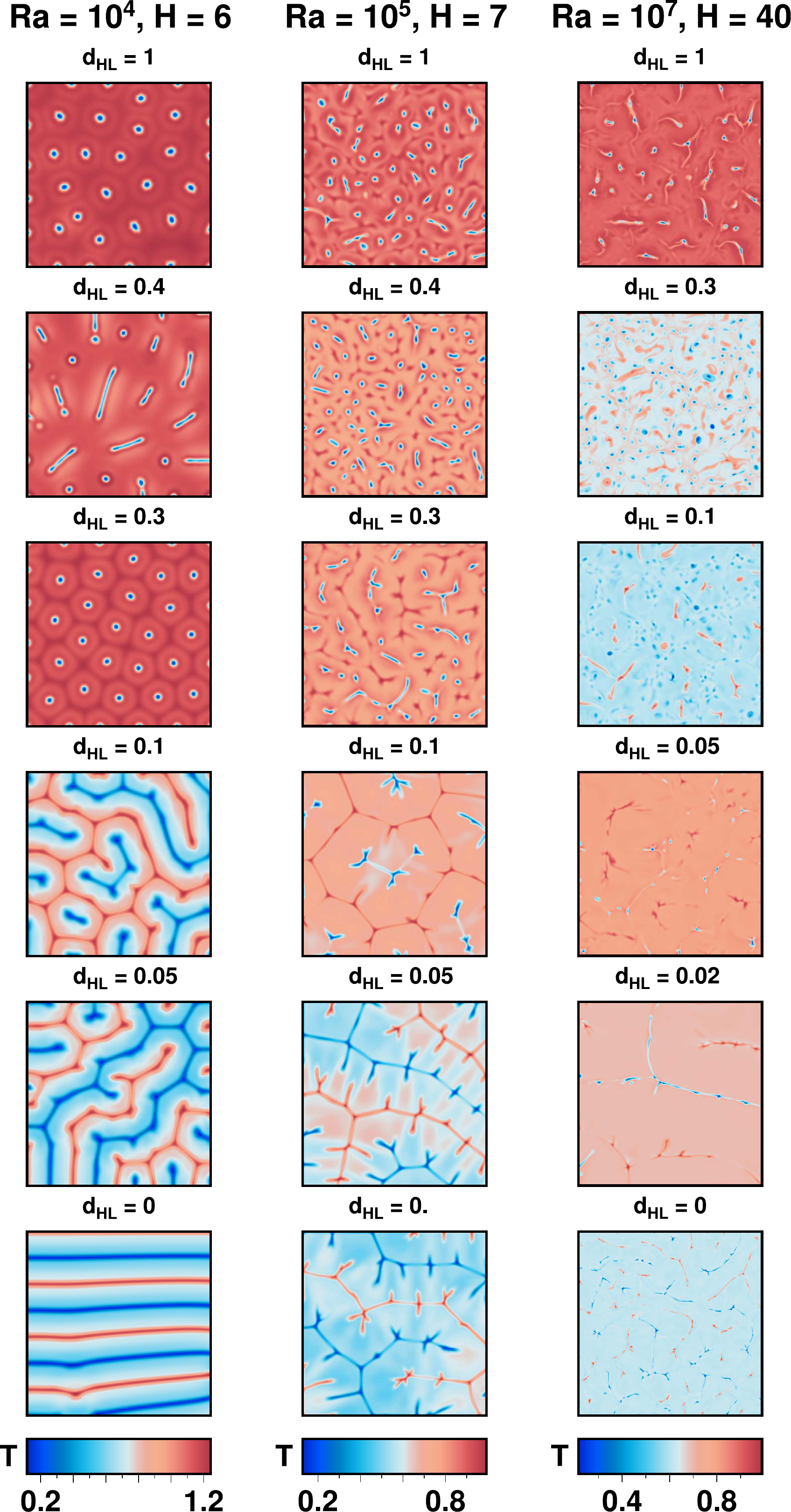}
\end{center}
\caption{\label{Planform} Convection planform at mid-depth for some cases described in table~\ref{run}. The colour scale changes for each case in order to enhance the visibility of the convective structure. The domain aspect ratio of the planform represented is kept constant at a given Rayleigh number corresponding to the lower aspect ratio available (table~\ref{run}).
}
\end{figure}

\begin{figure}
\begin{center}
\includegraphics[width=0.95\textwidth]{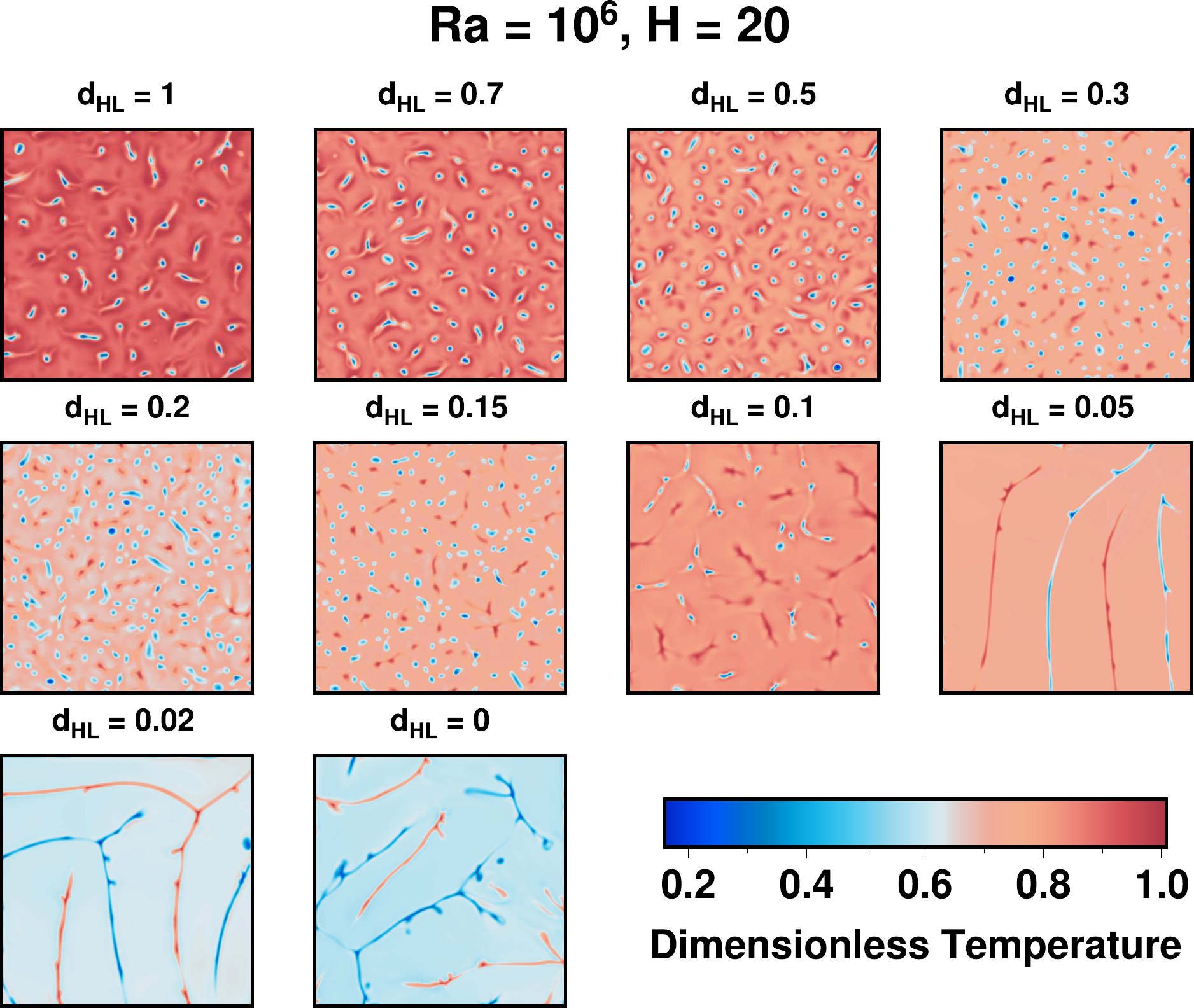}
\end{center}
\caption{\label{Planform_2} Convection planform at mid-depth for cases with $Ra = 10^{6}$ and $H = 20$ described in table~\ref{run}. The domain aspect ratio of the planform represented is kept constant to the lower aspect ratio available (table~\ref{run}).
}
\end{figure}

We have investigated the effects of heterogeneous internal heating by studying a convective system where internal heating is only present in a top horizontal layer. 
Different behaviours have been observed following the thickness of the heat blanket ($d_{HL}$) compared to the thickness of the thermal boundary layer in the homogeneous case ($\delta_{TBL}$).
For a ``thick'' heat blanket ($d_{HL} > \delta_{TBL}$), the convective system is generally equivalent to the system with homogeneous heating ($d_{HL} = 1.0$).
The only noteworthy effect is a decrease of the interior temperature with decreasing value of $d_{HL}$ (figures~\ref{HL_effects}b and~\ref{TProfil}a,b).
The case $d_{HL} \approx \delta_{TBL}$ marks a transition between two different regimes.
A peculiarity of this case is the peaked temperature observed in both the horizontally averaged and ``hot'' temperature profiles (figures~\ref{HL_effects}b and~\ref{TProfil}).
We interpret this feature as the result of a weakening of the generation of cold downwellings, which in turn reduces their ability to cool down the system.
For $d_{HL} < \delta_{TBL}$, all the system properties change quickly with decreasing value of $d_{HL}$ until it becomes equal to those obtained for the case without any internal heating ($d_{HL} = 0$).
In particular, the surface heat flux (figure~\ref{HL_effects}a) and convective vigour (figure~\ref{HL_effects}c) strongly increase with decreasing $d_{HL}$.
This can be explained by the appearance and strengthening of hot upwellings (figures~\ref{Planform} and~\ref{Planform_2}).
At the same time, the system is cooling down and tends to reach the thermal structure of the system without internal heating (figures~\ref{HL_effects}b and~\ref{TProfil}).
For $d_{HL} \ll \delta_{TBL}$, the convection planform does not change significantly as $d_{HL}$ decreases (figures~\ref{Planform} and~\ref{Planform_2}), while the surface heat flux and convective vigour are still modified substantially with decreasing $d_{HL}$.

Our results therefore suggest that there is a continuous change from the homogeneous heating case to the case without internal heating.
Nevertheless, an extremely thin heated layer is required to totally suppress the effect of the heated layer on the properties of the convective system.
As a consequence, the crustal enrichment in radioactive isotopes is a major feature of planetary bodies that should not be neglected.
For instance, the low surface heat flux observed beneath continents may be at least partly the result of such an enrichment.
Moreover, the small variations of the surface heat flux induced by the presence of the heat blanket support a minor impact of the presence of continents on the cooling rate of Earth.
By contrast, we speculate that in smaller rocky planets the cooling rate and the convective vigour may be reduced by the presence of an enriched crust, provided that the crustal thickness is slightly larger than $\delta_{TBL}$.
From a more general point of view, our results indicate the existence of different regimes characterized by very different properties depending on the crustal thickness.
These regimes may induce a certain variability of dynamics style on rocky planets.
The vertical distribution of heat producing elements would then be a key ingredient to understand the thermal state and evolution of a planet.
\section*{Acknowledgements}
We are grateful to Tobias Rolf and to another colleague for their careful and constructive reviews that helped us to greatly improve this manuscript.
This research was funded by the Ministry of Science and Technology of Taiwan (MOST) Grants 106-2116-M-001-014 and 107-2116-M-001-010 and by JSPS KAKENHI Grant JP19F19023.
Numerical computations were performed on IESAS Linux Cluster and on the cluster of Hokkaido University.
The interpretation of the results benefited from discussions with Claude Jaupart.
The data used for generating the figures are available for academic purposes \cite{Vilella2020c}.
The code used in this work is not publicly available but was thoroughly described in \citeA{Tackley2008}.

\clearpage
\begin{table}
{ \scriptsize
\begin{tabular}{l c c c c c c c c c c}
  & $d_{HL}$  & Resolution & Aspect & $\phi$ & $Ur$ & $\Delta \!T_{TBL}$ & $T_{1/2}$ & $V_{h}$ & $V_{rms}$ \\
  &           &            & ratio  &        &              &                    &           &         &    \\
          & 1    & 512 $\times$ 512 $\times$ 64    & 16:16:1 & 7.22$\pm$0.00    & 0.831 & 1.282 & 1.156 & 39.1 & 31.2\\  
          & 0.4  & 512 $\times$ 512 $\times$ 64    & 16:16:1 & 7.19$\pm$0.03    & 0.834 & 1.175 & 0.982 & 24.9 & 19.8\\
$Ra=10^4$ & 0.3  & 512 $\times$ 512 $\times$ 64    & 16:16:1 & 7.51$\pm$0.00    & 0.799 & 1.057 & 0.913 & 27.2 & 21.9\\
$H=6$     & 0.1  & 512 $\times$ 512 $\times$ 64    & 16:16:1 & 8.79$\pm$0.00    & 0.683 & 0.947 & 0.639 & 50.0 & 38.5\\  
          & 0.05 & 512 $\times$ 512 $\times$ 128   & 16:16:1 & 9.41$\pm$0.00    & 0.638 & 0.956 & 0.565 & 55.0 & 42.1\\
          & 0.   & 1024 $\times$ 1024 $\times$ 64  & 32:32:1 & 4.37$^*\pm$0.00  & 0.579 & 0.946 & 0.495 & 60.7 & 45.7\\
          &      &                                 &         &                  &       &       &       &      & \\
          & 1    & 768 $\times$ 768 $\times$ 128   & 12:12:1 & 12.43$\pm$0.09   & 0.563 & 0.979 & 0.850 & 111  & 103\\  
          & 0.4  & 768 $\times$ 768 $\times$ 128   & 12:12:1 & 12.06$\pm$0.07   & 0.580 & 0.969 & 0.777 & 86.7 & 79.5\\  
$Ra=10^5$ & 0.3  & 768 $\times$ 768 $\times$ 128   & 12:12:1 & 12.04$\pm$0.06   & 0.581 & 0.974 & 0.775 & 110  & 92.1\\                   
$H=7$     & 0.1  & 768 $\times$ 768 $\times$ 128   & 12:12:1 & 13.21$\pm$0.04   & 0.530 & 0.965 & 0.686 & 186  & 139\\  
          & 0.05 & 768 $\times$ 768 $\times$ 128   & 12:12:1 & 14.13$\pm$0.04   & 0.495 & 0.936 & 0.574 & 217  & 159\\
          & 0.   & 512 $\times$ 512 $\times$ 64    & 10:10:1 & 9.13$^*\pm$0.04  & 0.434 & 0.933 & 0.495 & 236  & 173\\
          &      &                                 &         &                  &       &       &       &      & \\
          & 1    & 512 $\times$ 512 $\times$ 128   & 8:8:1   & 28.66$\pm$0.26   & 0.698 & 1.023 & 0.912 & 338  & 276\\  
          & 0.7  & 768 $\times$ 768 $\times$ 192   & 8:8:1   & 28.38$\pm$0.19   & 0.705 & 1.029 & 0.871 & 268  & 214\\
          & 0.5  & 768 $\times$ 768 $\times$ 192   & 8:8:1   & 27.93$\pm$0.18   & 0.716 & 1.031 & 0.802 & 219  & 185\\  
          & 0.3  & 768 $\times$ 768 $\times$ 192   & 8:8:1   & 27.42$\pm$0.13   & 0.729 & 1.037 & 0.748 & 171  & 178\\    
$Ra=10^6$ & 0.2  & 768 $\times$ 768 $\times$ 192   & 8:8:1   & 27.12$\pm$0.12   & 0.737 & 1.034 & 0.733 & 159  & 195\\      
$H=20$    & 0.15 & 768 $\times$ 768 $\times$ 192   & 8:8:1   & 27.20$\pm$0.11   & 0.735 & 1.051 & 0.751 & 178  & 208\\                        
          & 0.1  & 768 $\times$ 768 $\times$ 192   & 8:8:1   & 27.61$\pm$0.19   & 0.724 & 0.989 & 0.799 & 330  & 295\\  
          & 0.05 & 1536 $\times$ 1536 $\times$ 384 & 12:12:1 & 30.67$\pm$0.11   & 0.652 & 0.977 & 0.707 & 662  & 505\\
          & 0.02 & 1536 $\times$ 1536 $\times$ 384 & 8:8:1   & 35.70$\pm$0.10   & 0.560 & 0.932 & 0.613 & 774  & 604\\
          & 0.   & 1024 $\times$ 1024 $\times$ 512 & 6:6:1   & 19.72$^*\pm$0.14 & 0.504 & 0.932 & 0.501 & 864  & 675\\
          &      &                                 &         &                  &                &       &       &      & \\
          & 1    & 768 $\times$ 768 $\times$ 192   & 8:8:1   & 57.95$\pm$0.41   & 0.690 & 0.996 & 0.882 & 1233 & 902\\  
          & 0.3  & 768 $\times$ 768 $\times$ 192   & 4:4:1   & 56.17$\pm$0.32   & 0.712 & 0.971 & 0.721 & 541  & 608\\
$Ra=10^7$ & 0.1  & 768 $\times$ 768 $\times$ 192   & 8:8:1   & 55.46$\pm$0.18   & 0.721 & 0.983 & 0.691 & 514  & 811\\ 
$H=40$    & 0.05 & 768 $\times$ 768 $\times$ 192   & 4:4:1   & 56.80$\pm$0.47   & 0.704 & 1.015 & 0.788 & 1253 & 1208\\
          & 0.02 & 1024 $\times$ 1024 $\times$ 384 & 6:6:1   & 66.35$\pm$0.44   & 0.603 & 0.967 & 0.690 & 2671 & 2138\\
          & 0.   & 1024 $\times$ 1024 $\times$ 512 & 8:8:1   & 41.45$^*\pm$0.47 & 0.491 & 0.927 & 0.500 & 2722 & 2364\\
\end{tabular}\\ 
$^*$ Note that the amount of internal heating $H$ should be added to the heat flux for a meaningful comparison with other cases.}
\caption{\label{run} Input parameters of the numerical simulations: the Rayleigh number ($Ra$), the dimensionless heating rate ($H$), the thickness of the heat blanket ($d_{HL}$), see text for more details, the grid resolution in X:Y:Z directions and the domain aspect ratio in the X:Y:Z directions. We also report some dimensionless characteristics of the system: $\phi$ the surface heat flux, $Ur = H/\phi$ the Urey ratio, $\Delta \!T_{TBL}$ the temperature jump across the top thermal boundary layer, $T_{1/2}$ the average temperature at mid-depth, $V_{h}$ the average surface velocity, $V_{rms}$ the volume average root mean square velocity.}
\end{table}
\clearpage
\clearpage
\clearpage
\begin{figure}
\begin{center}
\includegraphics[width=0.7\textwidth]{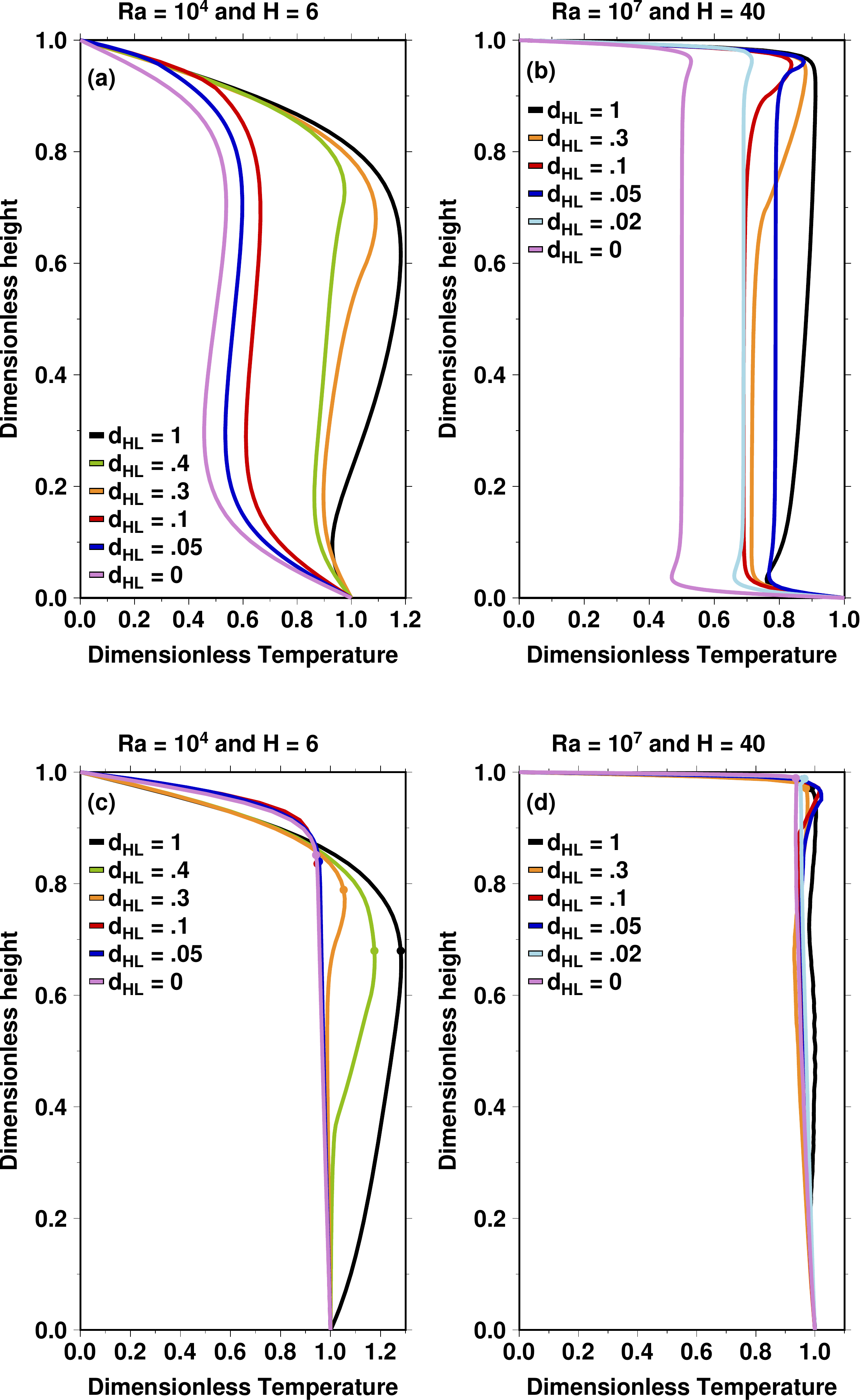}
\end{center}
\caption{\label{TProfil} Horizontally averaged temperature profiles (top panels) and ``hot'' temperature profiles (bottom panels) for numerical simulations conducted with (a, c) $Ra = 10^{4}$ with $H = 6$ and (b, d) $Ra = 10^{7}$ with $H = 40$. ``Hot'' temperature profiles are built from the hottest temperature at a given depth. The base of the thermal boundary layer is indicated by a circle (see Supplementary Material for more details).}
\end{figure}
\clearpage
\clearpage
\clearpage
\clearpage
\end{document}